\documentclass[hidelinks,aps,pra,preprint,superscriptaddress,10pt]{revtex4-2} 

\usepackage{amsmath,amssymb}
\usepackage{mathrsfs}
\usepackage{epstopdf}
\usepackage{graphicx}
\usepackage{bm,color,bbm}
\usepackage{natbib}
\usepackage{hyperref}
\usepackage{caption}
\usepackage{fancyhdr}
\usepackage{subcaption}
\newcommand{\nn}{\nonumber}

\begin{document}
\fancyhead[R]{\ifnum\value{page}<2\relax\else\thepage\fi}

\title{Nonlinear Optics as a source of high-dimensional genuine tripartite entanglement}

\author{James Schneeloch}
\email{james.schneeloch.1@us.af.mil}
\affiliation{Air Force Research Laboratory, Information Directorate, Rome, New York, 13441, USA}

\author{Richard J. Birrittella}
\affiliation{Air Force Research Laboratory, Information Directorate, Rome, New York, 13441, USA}

\author{Christopher C. Tison}
\affiliation{Air Force Research Laboratory, Information Directorate, Rome, New York, 13441, USA}

\author{Gregory A. Howland}
\affiliation{Microsystems Engineering, Rochester Institute of Technology, Rochester, New York 14623, USA}
\affiliation{School of Physics and Astronomy, Rochester Institute of Technology, Rochester, New York 14623, USA}

\author{Michael L. Fanto}
\affiliation{Air Force Research Laboratory, Information Directorate, Rome, New York, 13441, USA}

\author{Paul M. Alsing}
\affiliation{Air Force Research Laboratory, Information Directorate, Rome, New York, 13441, USA}


\date{\today}

\begin{abstract}
We lay down a general scheme to quantify the amount of genuine tripartite entanglement present in the spatial and energy-time degrees of freedom of entangled photon triplets using a resource-based measure known as the tripartite entanglement of formation. Quantifying genuine tripartite entanglement relative to a number of maximally entangled three-qubit Greenberger-Horne-Zeilinger (GHZ) states called gebits, the tripartite entanglement of formation serves as a basis of comparison between different tripartite entangled states of differing dimension. Demonstrating genuine tripartite entanglement is doubly challenging because it is not enough to show that each party is inseparable from the other two (which is sufficient only for pure states). Instead, one must rule out all mixtures of all combinations of biseparable states from describing the tripartite state.
To meet this challenge, we use entropic measures of the statistics of tripartite systems to simultaneously bound the correlations each party has with the other two, and in so doing place a lower limit to the tripartite entanglement of formation. Even though our measure never over-estimates the entanglement present, we estimate the effectiveness of our technique by determining the exact tripartite entanglement of a triple-gaussian triphoton wavefunction with the same correlations as seen in photon triplets generated in third-order spontaneous parametric down-conversion (SPDC). Between these two efforts, we show that a substantial amount of tripartite entanglement exists in both the spatial and energy-time degrees of freedom of these systems, and propose an experiment that can measure them.
\end{abstract}


\maketitle
\newpage
\thispagestyle{fancy}
\section{INTRODUCTION}
\label{sec:intro}  

Quantum entanglement is often cited as \emph{the} signature aspect of quantum physics that marks its departure from the classical regime, and with good reason. The exponentially large state space coming from entangled superpositions of multiple qubits is the basis for the advantages seen in quantum computing. The manifestly secure correlations from entanglement forms the bedrock of quantum networking. In addition, the enhanced sensitivity using entangled states of light allows researchers to probe beyond the standard quantum limit of shot noise. It is because of these applications that characterizing the \emph{amount} of entanglement present in a system as efficiently as possible is of critical importance toward benchmarking the utility and performance of new sources as they are being developed.

Historically, entanglement was originally considered a yes-or-no question. Over the last few decades \cite{PhysRevLett.80.2245,PhysRevLett.76.722,PhysRevA.57.1619,christandl2004squashed,PhysRevA.65.032314,RevModPhys.81.865}, many researchers have developed entanglement monotones and entanglement measures to compare different entangled states with one another, either by how far away they are from the set of separable (unentangled) states, or by its utility as a resource in entanglement-consuming tasks. While two-party entanglement has been well studied, and there are increasingly efficient ways of quantifying it, entanglement shared between three or more parties has comparably little research. In this work, we describe our strategy \cite{SchneelochQuantTripCVEnt2022} to quantify genuine tripartite entanglement in the high-dimensional degrees of freedom shared by entangled photon triplets. We describe the nonlinear-optical processes used to generate these triplets, and the properties of the triple-gaussian wavefunction describing their spatial and energy-time correlations.

\section{Quantifying tripartite entanglement}
Entanglement is defined as a departure from separability. Given three parties $A, B,$ and $C$, and a joint quantum state $\hat{\rho}_{ABC}$ shared by them, that state is considered separable either if the state factors out completely into $\hat{\rho}_{A}\otimes\hat{\rho}_{B}\otimes\hat{\rho}_{C}$, or if it can be expressed as a classical mixture of these separated states:
\begin{equation}
\hat{\rho}_{ABC}^{(sep)}=\sum_{i}p_{i}\Big(\hat{\rho}_{Ai}\otimes\hat{\rho}_{Bi}\otimes\hat{\rho}_{Ci}\Big).
\end{equation}
The correlations present in a separable state are exclusively classical, given the preparation probability distribution $\{p_{i}\}$. Any state that does not have this form is somehow entangled.

With only two parties, there is only one way a joint state can factor out as a product of individual states. With three parties, there are multiple forms of separability, which define multiple forms of entanglement. For example, states of the form $AB\otimes C$:
\begin{equation}
\hat{\rho}_{AB\otimes C}=\sum_{i}p_{i}\Big(\hat{\rho}_{ABi}\otimes\hat{\rho}_{Ci}\Big),
\end{equation}
are known as biseparable because they are expressible as mixtures of products of two states. For three parties, to be triseparable is to be fully separable.

Because there are multiple ways for a tripartite state to be in some way separable, demonstrating genuine tripartite entanglement is especially challenging. It is not enough to show that the state $\hat{\rho}_{ABC}$ is not within any class of biseparable or fully separable states. Instead, one must show that the state cannot be made using any ensemble of states coming from any of these classes of states.

To quantify genuine tripartite entanglement, we use a measure \cite{Szalay_MultiEntMeas} called the tripartite entanglement of formation $E_{3F}(ABC)$:
\begin{equation}
E_{3F}(ABC)=\min_{|\psi\rangle_{i}}\sum_{i}p_{i}\min\{S_{i}(A),S_{i}(B),S_{i}(C)\}.
\end{equation}
For each pure-state decomposition of $\hat{\rho}_{ABC}$, we find the minimum entanglement across all bipartitions (which would be zero for a biseparable state). Then, we find the minimum for this number across all possible pure-state decompositions of $\hat{\rho}_{ABC}$, which gives us the tripartite entanglement of formation $E_{3F}$. If $E_{3F}>0$, then there must be some genuinely tripartite-entangled state in every possible pure-state decomposition, proving the state is genuinely tripartite entangled.

The tripartite entanglement of formation $E_{3F}$ is a generalization of the ordinary entanglement of formation defined between two parties \cite{PhysRevLett.80.2245}:
\begin{equation}
E_{F}(AB)=\min_{|\psi\rangle_{i}}\sum_{i}p_{i}\min\{S_{i}(A),S_{i}(B)\}.
\end{equation}
For the ordinary entanglement of formation $E_{F}$, there is a standard unit of entanglement known as the \emph{ebit}, which is a two-qubit maximally entangled Bell state:
\begin{equation}
|\psi\rangle^{(ebit)}=\frac{1}{\sqrt{2}}\Big(|0,0\rangle +|1,1\rangle\Big).
\end{equation}
Using the entanglement of formation $E_{F}$, different two-party states can be compared relative to the number of ebits of equal value.

For the tripartite entanglement of formation $E_{3F}$, we define a unit of three-party entanglement known as the \emph{gebit}, which is the three-qubit Greenberger-Horne-Zeilinger (GHZ) state:
\begin{equation}
|\psi\rangle^{(gebit)}=\frac{1}{\sqrt{2}}\Big(|0,0,0\rangle +|1,1,1\rangle\Big).
\end{equation}
Similar to the ordinary entanglement of formation, the tripartite entanglement of formation $E_{3F}$ can be used to compare different three-party states relative to a corresponding number of gebits.  We use this purely as a side-by side comparison of different tripartite entangled states, as there are unresolved questions in interpreting this measure as a resource measure similar to the ordinary entanglement of formation.

There are complexities that arise in considering the gebit as a unit of tripartite entanglement. In particular, the $W$ state 
\begin{equation}
|W\rangle = \frac{1}{\sqrt{3}}\Big(|1,0,0\rangle + |0,1,0\rangle + |0,0,1\rangle \Big),
\end{equation}
is genuinely tripartite entangled, but cannot be expressed in terms of gebits. When synthesizing copies of genuinely multipartite entangled states using local operations and classical communication (LOCC), there is still no known set of resource states that can be used to accomplish this \cite{leuchs2019quantum}. Moreover, the distillable tripartite entanglement can actually be greater than the tripartite entanglement of formation because it is possible in principle to distill genuine tripartite entanglement from multiple copies of biseparable states as a resource \cite{guhne2009entanglement,PhysRevA.83.062321}. While there remain fundamental questions to be answered on the nature of tripartite entanglement as a resource, we use the tripartite entanglement of formation $E_{3F}$ to compare the amount of entanglement in an unknown quantum state to a corresponding number of gebits.

\subsection{Quantifying entanglement with correlations}
It is often said that entanglement is a kind of correlation. While correlations can exist without entanglement (i.e., classical correlations), entanglement cannot exist without correlations. Completely uncorrelated product states are separable by construction. Moreover, if the classical mutual information is zero in every local measurement basis, then the state is again completely uncorrelated \cite{ModiDiscordRMP2012}. Together, this implies that all entangled states must have some kind of measurable correlation.

Consider the following entanglement-correlation relation:
\begin{equation}\label{EntCorrRel}
H(Q_{A}:Q_{B})\leq E_{F}(AB) + \min\{S(AB),S(A),S(B)\}.
\end{equation}
See proof in Appendix \ref{AppCorRel}. Here, classical correlations are quantified by the Shannon mutual information $H(Q_{A}:Q_{B})$ of the joint probability distribution for the measurement outcomes of observables $\hat{Q}_{A}$ and $\hat{Q}_{B}$; the mixedness of the state $\hat{\rho}_{AB}$ is quantified by the joint von Neumann entropy $S(AB)$; and the mixedness of the subsystems $A$ and $B$ are quantified by their respective von Neumann entropies. For pure states of $AB$, ($S(AB)=0$), and all measurable correlations are identified as entanglement. For separable states, the correlations are bounded by the mixedness of the subsystems.

Because two-party entanglement is so closely identified with two-party correlations, we examined how this relationship works for tripartite entanglement.

In \cite{Schneeloch_TriEnt}, we showed how one could use the correlations present between three $D$-level quantum systems to begin to quantify the tripartite entanglement present. In particular, we found the relation:
\begin{equation}\label{OurDiscTripEntBound}
E_{3F}(ABC)\geq -S(A|BC)-S(B|AC)-S(C|AB)-2\log(D_{max}),
\end{equation}
where $D_{max}$ is the maximum dimension between systems $A$, $B$, and $C$; and where $S(A|BC)$ (for example) is the quantum conditional entropy equal to $S(ABC)-S(BC)$. For discrete observables $(Q_{A},Q_{B},Q_{C})$ and $(R_{A},R_{B},R_{C})$, one can use a version of the uncertainty principle in the presence of quantum memory \cite{Berta2010} to show:
\begin{equation}\label{BertaRel}
-S(A|BC)\geq\log(\Omega) -(H(Q_{A}|Q_{B},Q_{C})+H(R_{A}|R_{B},R_{C})),
\end{equation}
where
\begin{equation}
\Omega\equiv\min_{i,j}\left(\frac{1}{|\langle q_{i}|r_{j}\rangle|^{2}}\right),
\end{equation}
is a value that equals zero if observables $\hat{Q}_{A}$ and $\hat{R}_{A}$ commute, and approaches the dimension of system $A$, $D_{A}$ as $\hat{Q}_{A}$ and $\hat{R}_{A}$ approach being maximally uncertain with respect to one another (i.e., where the inner product of all eigenstates $|q_{i}\rangle$ of $\hat{Q}_{A}$ and $|r_{j}\rangle$ of $\hat{R}_{A}$ are equal to $1/\sqrt{D_{A}}$). This relation \eqref{BertaRel}, when substituted into our relation \eqref{OurDiscTripEntBound} allows us to quantify genuine tripartite entanglement in $D$-level systems. Because this strategy explicitly includes the dimensions of the systems, it could not be used to quantify entanglement in continuous-variable systems unless we make the a-priori assumption that the tripartite state is pure, which is experimentally unrealistic.

To solve this challenge, we developed a continuous-variable version of the uncertainty relation with quantum memory \cite{SchneelochPra2018}:
\begin{equation}
-S(A|BC)\geq \log(2\pi) - h(x_{A}|x_{B},x_{C}) - h(k_{A}|k_{B},k_{C}),
\end{equation}
where $h(x)$ is the continuous Shannon entropy \cite{Cover2006} of probability density $\rho(x)$, to create a corresponding entanglement bound for continuous variables \cite{SchneelochQuantTripCVEnt2022} where:
\begin{align}\label{OurContinuousRelation}
E_{3F}(ABC)&\geq \log(2\pi|\bar{\eta}||\bar{\beta|})-h(\eta_{A}x_{A} + \eta_{B}x_{B} + \eta_{C}x_{C})\nn\\
&\qquad\qquad-h(\beta_{A}k_{A} + \beta_{B}k_{B} + \beta_{C}k_{C}).
\end{align}
Here, $|\bar{\eta}||\bar{\beta}|=\min_{i}\{|\eta_{i}||\beta_{i}|\}$, and the coefficients $(\eta_{A},\eta_{B},\eta_{C})$ and $(\beta_{A},\beta_{B},\beta_{C})$ may be adjusted to best fit the correlations naturally present in the system.

\section{Entangled photon triplets in nonlinear optics}
Nonlinear optics has been the preferred source for entangled photon pairs for decades. In short, photons from a pump laser pass through a transparent medium, and the atoms within that medium respond to the oscillating electric field. When this disturbance is small enough that the response of the medium is in proportion to the electric field, we are in the domain of linear optics. Nonlinear optics exists in the regime where the field is strong enough that the atoms respond anharmonically, but before the electrons are ripped off of the atoms altogether.

The response of the material in nonlinear optics is described as a power series of the induced polarization of light $\vec{P}$ as a function of the electric field $\vec{E}$ \cite{boyd2007nonlinear}:
\begin{equation}
P_{i}=\epsilon_{0}\Big(\chi_{ij}^{(1)}E_{j} + \chi_{ijk}^{(2)}E_{j}E_{k}+ \chi_{ijk\ell}^{(3)}E_{j}E_{k}E_{\ell}+...\Big).
\end{equation}
Here, $\chi_{ij}^{(1)}$ gives the linear optical susceptibility, while $\chi_{ijk}^{(2)}$ and $\chi_{ijk\ell}^{(3)}$ give the second and third-order nonlinear optical susceptibilities respectively. 

Just as in acoustic vibrations, an applied electric field of a single frequency can induce the generation of light at harmonics of that electric field. What determines how much of each harmonic gets generated is in the magnitude of the nonlinear susceptibilities, and whether the dispersion in the material makes momentum conservation possible (where the index of refraction affects the wavelength and therefore the momentum of photons).

Quantum mechanically, generating light at a frequency twice that of the original pump beam requires pairs of pump photons to be consumed to generate each photon at the second harmonic. Since the fundamental conditions that optimize second-harmonic generation are only concerned with total momentum conservation and the scale of the susceptibility, they are also the conditions that optimize the reverse process where pairs of photons (here called the signal and idler photons) are generated at frequencies adding up to the frequency of an annihilated pump photon. This $\chi^{(2)}$ process, known as Spontaneous Parametric Down-Conversion (SPDC), has been the most popular source of entangled photon pairs in quantum information for decades. Where the emitted biphotons are regarded as being nearly in a pure state, the strong position correlations that come from the pair having a common point of origin (the birth zone \cite{Schneeloch_SPDC_Reference_2016}), and the strong momentum correlations coming from momentum conservation with the original pump photon imply a large degree of spatial entanglement can be created \cite{SchneelochPra2018}. Similarly, from a common time of generation, and energy conservation, a large amount of energy-time entanglement can be generated between these photon pairs as well.

To create entangled photon triplets in nonlinear optics, there are two major approaches. The first is to use a $\chi^{(3)}$ process that is the third-order analogue of SPDC, and the reciprocal process to third-harmonic generation. In this case, photon triplets are created in single quantum events from each annihilated pump photon. The second approach is to use $\chi^{(2)}$-SPDC twice; first to split the pump photon into signal and idler, and the second to split the signal photon into two again to yield three photons in total. In both approaches, momentum conservation and common places of origin give rise to correlations and entanglement. From this, the coefficients $(\eta_{A},\eta_{B},\eta_{C})=(1,-1/2,-1/2)$ and $(\beta_{A},\beta_{B},\beta_{C})=(1,1,1)$ will allow our tripartite entanglement relation to take maximum advantage of these correlations.

Since both approaches to generate entangled photon triplets cover the same kind of correlations, we consider the example of third-order $\chi^{(3)}$-SPDC. The case of cascaded $\chi^{(2)}$-SPDC is covered in \cite{SchneelochQuantTripCVEnt2022}.

In $\chi^{(3)}$-SPDC, the triphoton wavefunction in one transverse dimension has the form:
\begin{align}
\psi(&k_{1},k_{2},k_{3})\approx\mathcal{N} \alpha_{p}(k_{1}+k_{2}+k_{3})\\
&\times \text{Sinc }\bigg(\Big(|k_{2}+k_{3}|^{2}+|k_{1}+k_{3}|^{2}+|k_{1}+k_{2}|^{2}\Big)\frac{3L_{z}}{4|\vec{k}_{p}|}\bigg),\nn
\end{align}
where $\alpha_{p}(k_{1}+k_{2}+k_{3})$ is the transverse pump momentum amplitude, which with momentum conservation becomes a function of $k_{1}+k_{2}+k_{3}$. See Appendix \ref{AppTriphotDeriv} for details. Like in ordinary SPDC, the sinc function coming from integrating the phase matching exponential $e^{i\Delta\vec{k}\cdot\vec{r}}$ over the dimensions of the nonlinear medium, except here we have a different momentum mismatch $\Delta \vec{k}\equiv \vec{k}_{1}+\vec{k}_{2}+\vec{k}_{3}-\vec{k}_{p}$.

Our expression for $\psi(k_{1},k_{2},k_{3})$, while complicated, greatly simplifies if we use the rotated coordinates:
\begin{subequations}
\begin{align}
k_{u}&=\frac{1}{\sqrt{3}}(k_{1}+k_{2}+k_{3}),\\
k_{v}&= \sqrt{\frac{2}{3}}\Big(-k_{1} +\frac{k_{2}+k_{3}}{2}\Big),\\
k_{w}&=\frac{1}{\sqrt{2}}(k_{2}-k_{3}),
\end{align}
\end{subequations}
so that
\begin{equation}
\psi(k_{u},k_{v},k_{w})\approx\mathcal{N} \alpha_{p}(k_{u}\sqrt{3}) \text{Sinc }\bigg(\frac{3L_{z}}{4|\vec{k}_{p}|}\Big(4 k_{u}^{2} + k_{v}^{2} + k_{w}^{2}\Big)\bigg).
\end{equation}
Just as the biphoton wavefunction in ordinary SPDC for a gaussian pump beam can be approximated as a double-Gaussian wavefunction, the triphoton wavefunction can be approximated as a triple-gaussian wavefunction:
 \begin{equation}
\psi(k_{u},k_{v},k_{w})\approx\mathcal{N}e^{-(\frac{32a}{9}+3\sigma_{p}^{2})k_{u}^{2}}e^{-\frac{8a}{9}(k_{v}^{2}+k_{w}^{2})}
 \end{equation}
 where $a=\frac{3L_{z}}{4 k_{p}}$ and $\sigma_{p}$ is the pump beam radius in position space (such that the $1/e^{2}$ beam diameter is $4\sigma_{p}$).

In a similar fashion, one may derive a corresponding wavefunction for the energy/time correlations of the entangled triphotons from its joint spectral amplitude. For a narrow bandwidth pump, the joint spectral amplitude has the form:
\begin{equation}
\psi(\delta\omega_{1},\delta\omega_{2},\delta\omega_{3})=\mathcal{N'}s(\sqrt{3}\delta\omega_{u})\text{Sinc}(\frac{\kappa_{0}L_{z}}{4}(\delta\omega_{v}^{2}+\delta\omega_{w}^{2}))
\end{equation}
where $\kappa_{0}$ is the group-velocity dispersion at the central frequency of the down-converted photon triplet (i.e., $|d^{2}k/d\omega^{2}|_{\omega_{10}}$), and where for example $\delta\omega_{1}=\omega_{1}-\omega_{10}$ is the detuning of $\omega_{1}$ form its central frequency, and $(\delta\omega_{u},\delta\omega_{v},\delta\omega_{w})$ are rotated frequency coordinates from $(\delta\omega_{1},\delta\omega_{2},\delta\omega_{3})$ similar to those just used in momentum. Here, $s(\delta\omega_{p})$ is the pump spectral amplitude (relative to pump central frequency) where energy conservation allows us to make the substitution $\delta\omega_{p}=\delta\omega_{1}+\delta\omega_{2}+\delta\omega_{3}=\sqrt{3}\delta\omega_{u}$.  See Appendix \ref{AppTriPhotBrightness} for derivation and details. Since these energy-time tripartite correlations are of the same quality as the spatial correlations, we focus on the spatial correlations for the remainder of this work.

\section{The Triple-Gaussian wavefunction and tripartite entanglement}
To describe the tripartite entanglement available in entangled photon triplets, we will first look at the triple-gaussian wavefunction that fits these correlations, and observe its entanglement properties.

For two-party entanglement in continuous-variable wavefunctions, the double-Gaussian wavefunction is an approximate generalization of the Bell state to high dimension:
\begin{equation}
\psi(x_{1},x_{2})=\frac{1}{\sqrt{2\pi\sigma_{+}\sigma_{-}}}e^{-\frac{(x_{1}+x_{2})^{2}}{8\sigma_{+}^{2}}}e^{-\frac{(x_{1}-x_{2})^{2}}{8\sigma_{-}^{2}}},
\end{equation}
which in the rotated coordinates:
\begin{equation}
x_{+}\equiv \frac{x_{1}+x_{2}}{\sqrt{2}}\qquad:\qquad x_{-}\equiv \frac{x_{1}-x_{2}}{\sqrt{2}},
\end{equation}
separates into the product:
\begin{equation}
\psi(x_{+},x_{-})=\frac{1}{(2\pi\sigma_{+}^{2})^{1/4}}e^{-\frac{x_{+}^{2}}{4\sigma_{+}^{2}}}\frac{1}{(2\pi\sigma_{-}^{2})^{1/4}}e^{-\frac{x_{-}^{2}}{4\sigma_{-}^{2}}}.
\end{equation}
Here, $\sigma(x_{1}+x_{2})=\sigma_{+}\sqrt{2}$ and $\sigma(x_{1}-x_{2})=\sigma_{-}\sqrt{2}$. For a biphoton sharing a common point of origin in an otherwise large beam, $\sigma_{-}\ll\sigma_{+}$, and there will be strong correlations and entanglement. In the limit that $\sigma_{-}$ approaches zero, we approach a generalization of the Bell state to continuous variables known as the EPR wavefunction \cite{EPR1935}. Because of how position and momentum space are related by fourier transforms, we can transform the double-gaussian function readily to momentum space by the substitution:
\begin{equation}
\sigma_{k_{+}}=\frac{1}{2\sigma_{+}}\qquad:\qquad \sigma_{k_{-}}=\frac{1}{2\sigma_{-}}.
\end{equation}
From this, we can also see that the same regime that provides strong position correlations ($\sigma_{+}\gg\sigma_{-}$) will also provide strong momentum anti-correlations ($\sigma_{k_{-}}\gg\sigma_{k_{+}}$).

For three-party entanglement in continuous-variable wavefunctions, we have the triple-gaussian wavefunction, which is an approximate generalization of the GHZ state:
\begin{equation}
\psi(x_{1},x_{2},x_{3}) =\frac{e^{-\frac{(x_{1}+x_{2}+x_{3})^{2}}{12\sigma_{u}^{2}}- \frac{\left(x_{1}-\frac{x_{2}+x_{3}}{2}\right)^{2}}{6\sigma_{v}^{2}}-\frac{(x_{2}-x_{3})^{2}}{8\sigma_{w}^{2}}}}{\sqrt{(2\pi)^{3/2} \sigma_{u}\sigma_{v}\sigma_{w}}},
\end{equation}
The triple-gaussian wavefunction (when $\sigma_{v}=\sigma_{w}$) describes the spatial correlations of entangled photon triplets that are symmetric under permutation of parties. In general, the momentum bandwidth of the pump beam, plus momentum conservation (or alternatively just the pump beam diameter) mostly determines $\sigma_{u}$, while only phase matching determines $\sigma_{v}$ (and therefore $\sigma_{w}$).

When expressed in the rotated coordinates:
\begin{subequations}
\begin{align}
x_{u}&=\frac{1}{\sqrt{3}}(x_{1}+x_{2}+x_{3}),\\
x_{v}&=\frac{2}{\sqrt{6}}\left(-x_{1}+\frac{x_{2}+x_{3}}{2}\right),\\
x_{w}&=\frac{1}{\sqrt{2}}(x_{2}-x_{3}),
\end{align}
\end{subequations}
the triple-Gaussian wavefunction simplifies substantially to:
\begin{equation}
\psi(x_{u},x_{v},x_{w}) =\frac{e^{-\frac{x_{u}^{2}}{4\sigma_{u}^{2}}}e^{-\frac{x_{v}^{2}}{4\sigma_{v}^{2}}}e^{-\frac{x_{w}^{2}}{4\sigma_{w}^{2}}}}{\sqrt{(2\pi)^{3/2} \sigma_{u}\sigma_{v}\sigma_{w}}}.
\end{equation}
Here we can appreciate that the triple-gaussian wavefunction of entangled photon triplets is a gaussian spheroid that transforms from a pancake-shaped distribution in momentum space (momenta nearly constrained to the plane $k_{1}+k_{2}+k_{3}\approx const$) to a sausage-shaped distribution in position space (position of one determines the other two to within a birth zone). As defined for biphotons \cite{Schneeloch_SPDC_Reference_2016}, the triphoton birth zone $\Delta_{BZ}$ here is twice the standard deviation of the initial position of one of the down-converted photons $x_{1}$ conditioned on the nominal position of the annihilated pump photon $x_{p}=(x_{1}+x_{2}+x_{3})/3$. With this definition, we get:
\begin{equation}
\Delta_{BZ}=2\sigma(x_{1}|x_{p})\approx\sqrt{\frac{4}{3}}\sigma(x_{1}-x_{2})=\frac{4}{3}\sigma\left(x_{1}-\frac{x_{2}+x_{3}}{2}\right),
\end{equation}
where the approximation comes from the triple-gaussian case where the distribution factors into a product of distributions for $x_{u}$, $x_{v}$ and $x_{w}$, respectively. In the case of entangled photon triplets generated in multiple-stage processes (such as cascaded $\chi^{2}$-SPDC), the triphoton birth zone is more a measure of photon position correlations rather than the size of the region in which the triphoton was approximately generated.

Remarkably, the tripartite entanglement of formation can be calculated exactly for this triple gaussian wavefunction as discussed in \cite{SchneelochQuantTripCVEnt2022}:
\begin{equation}
E_{3F}(ABC)=\frac{h_{2}(\lambda_{0})}{\lambda_{0}},
\end{equation}
where $h_{2}(\lambda_{0})$ is the binary entropy function:
\begin{equation}
h_{2}(\lambda_{0})=-\lambda_{0}\log_{2}(\lambda_{0})-(1-\lambda_{0})\log_{2}(1-\lambda_{0}),
\end{equation}
and 
\begin{equation}
\lambda_{0}=\frac{2}{1+\frac{1}{3}\sqrt{5+2\left(\frac{\sigma_{u}^{2}}{\sigma_{v}^{2}} + \frac{\sigma_{v}^{2}}{\sigma_{u}^{2}}\right)}}.
\end{equation}
Here we see that whenever the triple-Gaussian state is not fully separable (i.e., $\sigma_{u}\neq\sigma_{v}$), there is a nonzero amount of tripartite entanglement. Moreover, this tripartite entanglement may be made arbitrarily large when either $\sigma_{u}\gg\sigma_{v}$ or $\sigma_{v}\gg\sigma_{u}$.

The GHZ state possesses no two-party entanglement. Although the triple Gaussian wavefunction is an approximate generalization of the GHZ state to high dimension, one can show that unless fully separable, it always has some small amount of two-party entanglement. To show this, we look at the correlation statistics of a two-party subsystem of the tripartite triple-Gaussian state.

The marginal position probability distribution of the triple Gaussian wavefunction is a double-Gaussian:
\begin{equation}
\rho(x_{B},x_{C})=\frac{1}{2\pi\sigma_{u}\sigma_{v}^{2}\sqrt{\frac{1}{3\sigma_{u}^{2}} +\frac{2}{3\sigma_{v}^{2}}}}e^{-\frac{3}{4(2\sigma_{u}^{2} + \sigma_{v}^{2})}(x_{B}+x_{C})^{2}}e^{-\frac{(x_{B}-x_{C})^{2}}{4\sigma_{v}^{2}}},
\end{equation}
from which we have the statistics:
\begin{equation}
\sigma(x_{B}+x_{C})=\sqrt{\frac{4\sigma_{u}^{2}+2\sigma_{v}^{2}}{3}}\qquad:\qquad \sigma(x_{B}-x_{C})=\sigma_{v}\sqrt{2}.
\end{equation}

The marginal momentum probability distribution of the triple Gaussian wavefunction is also a double-Gaussian. Since the triple gaussian wavefunction factors when using separated coordinates, we know from the fourier transform relations of a gaussian wavefunction that:
\begin{equation}
\sigma_{ku}=\frac{1}{2\sigma_{u}}\qquad:\qquad \sigma_{kv}=\frac{1}{2\sigma_{v}}.
\end{equation}
In short, we may take the momentum-space triple-gaussian probability density to be identical to the position-space density, once these preceding substitutions are made.

In momentum space, we have the correlation statistics:
\begin{equation}
\sigma(k_{B}+k_{C})=\sqrt{\frac{1}{3\sigma_{u}^{2}} +\frac{1}{6\sigma_{v}^{2}}}\qquad:\qquad \sigma(k_{B}-k_{C})=\frac{1}{\sigma_{v}\sqrt{2}}.
\end{equation}

Here, we discover a tradeoff in correlations. To obtain strong position correlations, one must have $\sigma_{v}\ll\sigma_{u}$. However, to obtain strong momentum correlations one must be in the opposite regime where $\sigma_{v}\gg\sigma_{u}$. If there is any two-party entanglement in these two subsystems, it has a strict upper limit. Using these statistics, we can examine whether any entanglement can be witnessed using these correlations.

In 2002, Mancini \emph{et.~al} \cite{PhysRevLett.88.120401} developed a strong separability criterion that works for continuous Fourier-conjugate variables (e.g., position/momentum). In particular, all separable states must obey the condition 
\begin{align}
0&\geq-\log(\sigma(x_{B}\mp x_{C})\sigma(k_{B}\pm k_{C})),
\end{align}
which in terms of the triple-gaussian wavefunction becomes:
\begin{equation}
0\geq\max\left\{-\log\Big(\sqrt{\frac{2\sigma_{v}^{2}}{3\sigma_{u}^{2}} + \frac{1}{3}}\Big),-\log\Big(\sqrt{\frac{2\sigma_{u}^{2}}{3\sigma_{v}^{2}} + \frac{1}{3}}\Big)\right\}.
\end{equation}
For $\sigma_{u}=\sigma_{v}$ the triple gaussian state is fully separable, and the right-hand side of this criterion is zero. For all other cases, the right-hand side of this criterion exceeds zero, but has a small upper limit of about $+0.7925$ bits (unlike for tripartite entanglement, which has no upper limit). This can be understood through the entanglement correlation relation \eqref{EntCorrRel}. In the limit of large tripartite entanglement $S(C)$ is large, but so is $S(AB)$. This means that the measured correlations between $A$ and $B$ may be large without there needing to be large entanglement between $A$ and $B$. Still, this means that unlike a true GHZ state, the triple-Gaussian wavefunction is either fully separable, or else possesses both two- and three-party entanglement.

\section{Effectiveness of tripartite correlations at quantifying tripartite entanglement}
With the basics of tripartite entanglement of the triple-gaussian wavefunction covered, we can compare how well our entanglement quantification strategy  works compared to the approximate tripartite entanglement present.

Given the triple gaussian wavefunction
 \begin{equation}
\psi(k_{u},k_{v},k_{w})\approx\mathcal{N}e^{-(\frac{32a}{9}+3\sigma_{p}^{2})k_{u}^{2}}e^{-\frac{8a}{9}(k_{v}^{2}+k_{w}^{2})},
 \end{equation}
 we can express our tripartite entanglement criterion in terms of these coordinates:
 \begin{equation}
E_{3F}(ABC)\geq-\log(3\sqrt{2}e) -\frac{1}{2}\log(\sigma_{v}^{2}\sigma_{ku}^{2}),
 \end{equation}
 with variances
 \begin{equation}
\sigma_{v}^{2}=\frac{8a}{9}\qquad:\qquad \sigma_{ku}^{2}=\frac{1}{4\left(\frac{32a}{9} + 3\sigma_{p}^{2}\right)},
 \end{equation}
 to obtain our bound:
 \begin{equation}
 E_{3F}(ABC)\geq \frac{1}{2}\log\left(16 + \frac{18\sigma_{p}^{2}k_{p}}{L_{z}}\right)-\log(3\sqrt{2}e).
 \end{equation}

To put this bound into practice, let us input reasonable experimental parameters. Examples of the the reverse process (THG) are seen in \cite{tomov1992third,miyata2011high}. Consider a pump beam of wavelength of $516.67$nm incident on a $\chi^{(3)}$ nonlinear medium of length $L_{z}=3$mm. Let us consider degenerate $\chi^{(3)}$-SPDC into photon triplets of wavelength $1550$nm. For some materials, this phase matching may be achieved by angle tuning the medium and using its birefringence similar to how most $\chi^{(2)}$-SPDC is carried out in other bulk crystals. To expand the range of capability, it would be useful to achieve quasi-phase matching by periodically varying $\chi^{(3)}$ over the length of the medium. However, periodic poling does not flip the sign of $\chi^{(3)}$ as it does for $\chi^{(2)}$. Instead other strategies must be carried out such as periodically doping the medium, or in using the intensity-dependent refractive index along with an illumination that varies periodically along the medium. To phase-match PPLN, we consider type-I third-order SPDC, where the photon triplet has a polarization orthogonal to the pump (but equal to each other), and with a quasi-phase matching period of $17.69\mu$m using dispersion data from \cite{gayer2008temperature} .

\begin{figure}[ht!]
\includegraphics[width=0.6\columnwidth]{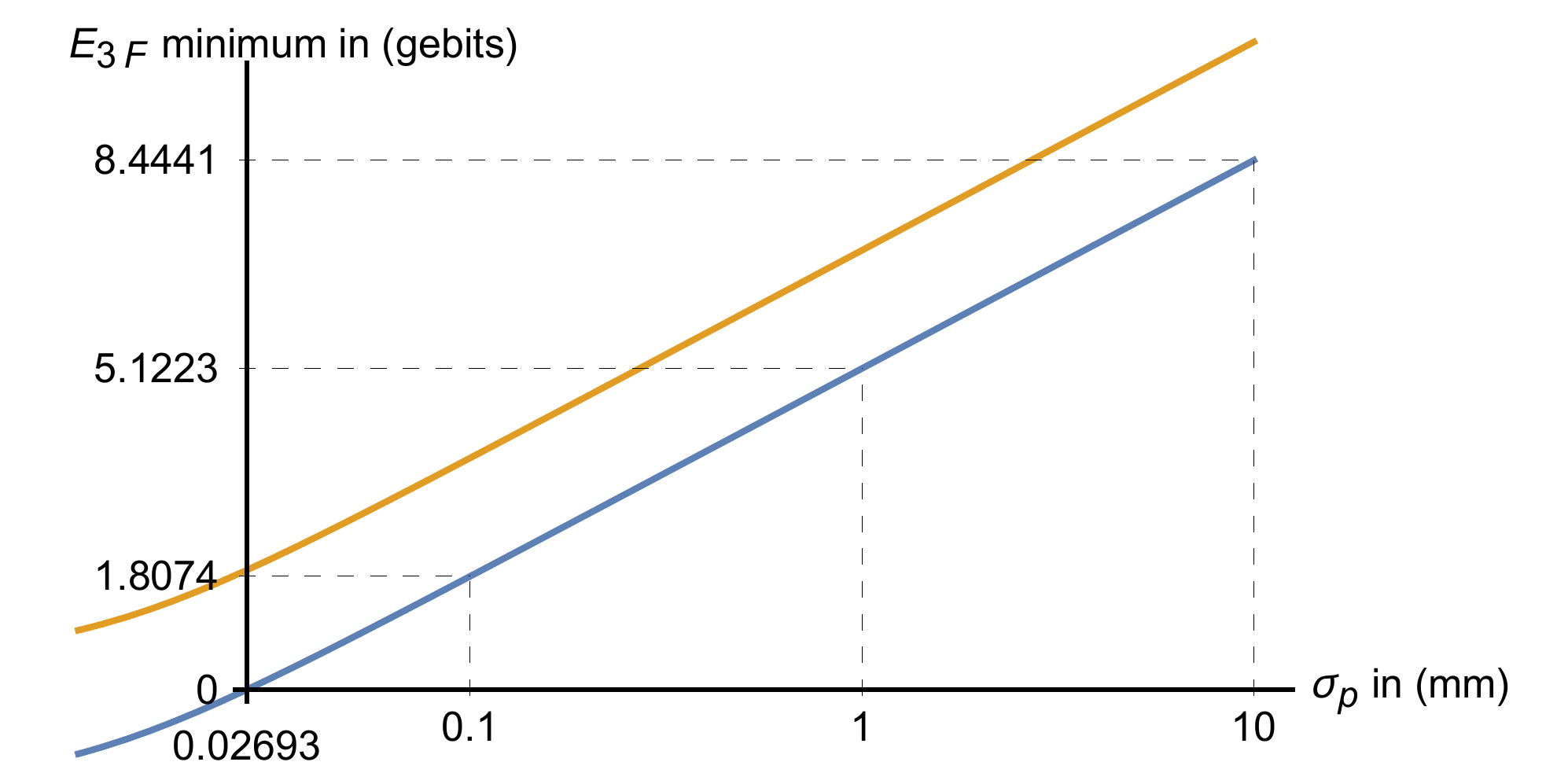}
\centering
\caption{Semi-log plot showing $E_{3F}$ as a function of pump beam radius $\sigma_{p}$ for the triphoton wavefunction in one transverse spatial degree of freedom used in third-order degenerate SPDC. The blue curve represents the amount of tripartite entanglement that can be witnessed with these correlations. The yellow curve above the blue curve represents the exact tripartite entanglement in the system assuming the triphoton wavefunction is well-described as a triple-gaussian wavefunction with the same correlations. In the limit of large $\sigma_{p}$, the two curves approach a constant offset of $1-2/\ln(2)$ or about $-1.8854$ gebits.}\label{CVEntPlot}
\end{figure}

With these experimental parameters $\lambda_{p}=516.67$nm and $L_{z}=3$mm, we have plotted in Fig.~\ref{CVEntPlot} the amount of spatial entanglement that can be witnessed as a function of the pump beam radius $\sigma_{p}$, as well as the approximate total entanglement assuming a triple-Gaussian wavefunction. Here we see that there can be a substantial amount of tripartite spatial entanglement within these systems (where for example $5$ gebits together constitutes $15$ qubits).

\section{Discussion: Proposed experiment and technical challenges}
\begin{figure*}[ht!]
\includegraphics[width=0.7\textwidth]{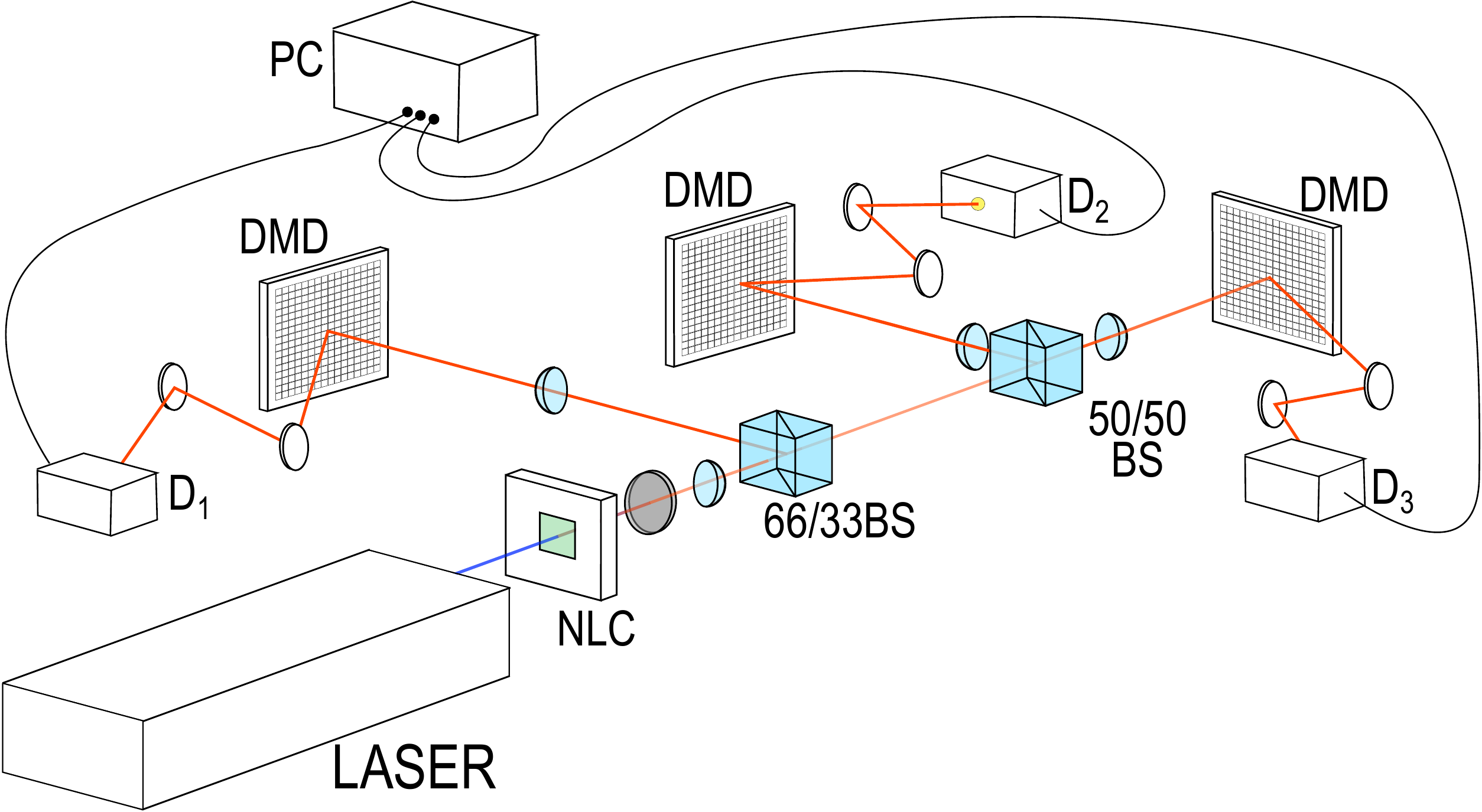}
\centering
\caption{Basic experimental diagram on using DMD arrays in concert with imaging optics and a Photon correlator (PC) to measure triplet coincidence counts from a nonlinear crystal (NLC) pumped by a laser, one can obtain the joint position probability distribution of the generated photon triplets.}\label{ExpDiagram}
\end{figure*}

Assuming we have a suitable source of tripartite entanglement, we can utilize the same kind of experiment used to measure two-party spatial entanglement in \cite{schneeloch2018EntExp}. In short, one images the face of the nonlinear medium (or its Fourier transform onto three digital micromirror device (DMD) arrays whose pixels can be directed toward or away from respective photon detectors (See Figure~\ref{ExpDiagram} for diagram). To accomplish this, the light is split first with a variable beamsplitter tuned to amplitudes $(\sqrt{1/3},\sqrt{2/3})$, where the spatial mode associated to the $\sqrt{1/3}$ amplitude is directed toward the first DMD. The light in the exiting spatial mode associated to the $\sqrt{2/3}$ amplitude is split again, this time with an ordinary $50/50$ beamsplitter, whose exiting modes are directed toward their respective DMD arrays. Using this pair of beamplitters maximizes the amplitude for the photon triplets to be separated, assuming they are indistinguishable in polarization/frequency. These detections will be correlated with photon correlators in order to obtain a photon triplet coincidence count rate. By examining how the triplet photon coincidence count rate changes as a function of the patterns we place on each DMD array, we may build up either the joint position probability distribution or the joint momentum distribution for the photon triplets, which may be used to conservatively estimate the spatial entanglement present.

To obtain these photon triplet statistics there are both technical and fundamental challenges that need to be addressed. The major technical challenge to this approach is in collection efficiency. If we only collect those photon triplets that reflect off of one triplet of micromirrors at a time, and scan over all possible pixel triplets, this collection will be completely impractical because of the large number of pixel triplets, and the small fraction of the total illumination that would correspond to a particular triplet. In \cite{schneeloch2018EntExp} , we solved this challenge for photon pair sources by using a smart multi-resolution sampling scheme. Instead of scanning over individual pixel triplets, we would first measure the spatial correlations at the lowest possible resolution (where say half of each DMD would be all-on or all-off). Then, were we detect significant coincidence counts, we would subdivide those regions, and measure them at twice the resolution. Those sub-regions with significant correlations would be further subdivided and sampled until we have obtained a coarse-grained approximation of the biphoton joint probability distribution that is sampled at high resolution only where there is significant probability, but without overlooking regions of low probability. Because our tripartite entanglement quantifier \eqref{OurContinuousRelation} is based on continuous entropy, and coarse-graining cannot decrease continuous entropy, we can use these same methods to quantify the tripartite entanglement present in these systems without over-estimating it.

The major fundamental challenges in quantifying high-dimensional tripartite entanglement are in efficiency and confounding secondary processes. While third-harmonic generation (THG) can be obtained with reasonably high efficiency at sufficiently high input powers, this works primarily because the efficiency of THG actually increases with increasing input power. The reverse process of $\chi^{(3)}$-SPDC is a process whose efficiency is relatively constant with pump power, and is low enough that demonstrations of photon triplet generation of more than ten triplets per minute \cite{hamel2014direct,PhysRevA.101.033840} are highly significant. Alternatively, one can consider a resonant system where the pump laser power may build up to a large value in a micro-ring resonator made out of a $\chi^{(3)}$ medium, but where the down-converted photon triplets are readily extracted.

For an estimate of the number of generated photon triplets in a $\chi^{(3)}$ medium, we assume a single-mode gaussian pump beam of power $P$ and radius $\sigma_{p}$; a medium of length $L_{z}$, and wide enough in both directions to encompass that beam, and photon triplets with equal focal parameter to the pump such that, say $\sigma_{1}=\sigma_{p}\sqrt{3}$. This could be used to describe triplet generation in a single-mode waveguide. With these assumptions (as well as assuming perfect phase matching), one can use similar methods to those employed in \cite{Schneeloch_2019_RateRef} to show:
\begin{equation}
\langle R_{triplets}\rangle = \frac{\hbar }{2592\sqrt{3}\pi^{2}\epsilon_{0}^{2}c^{4}}\frac{n_{g1}n_{g2}n_{g3}n_{gp}}{n^{2}_{p}n^{2}_{1}n^{2}_{2}n_{3}^{2}}\frac{(\chi_{eff}^{(3)})^{2}\omega_{p0}^{3}}{|\kappa_{0}|}\frac{P L_{z}}{\sigma_{p}^{4}}.
\end{equation}
where here, $\kappa_{0}$ is the group-velocity dispersion (second derivative of $k(\omega)$) at the central frequency of the down-converted photon triplets. See Appendix \ref{AppTriPhotBrightness} for abbreviated derivation. If we use statistics of fused silica \cite{malitson1965interspecimen,PhysRevB.61.10702}, estimated for third-order downconversion from $516.67\rightarrow 1550$nm (using Miller's Rule to interpolate between the values of $\chi^{(3)}_{eff}$ in \cite{PhysRevB.61.10702}), and assume a $10\mu$m $(1/e^{2})$ beam diameter for the down-converted spatial modes, we obtain the simplified formula:
\begin{equation}
R_{triplets}\approx (11.6267)L_{z}P
\end{equation}
for a $10$cm-long waveguide, one would need a relatively large $143mW$ of pump power to start generating in excess of ten triplets per minute. This prediction does not take into account the task of ensuring phase matching, which with this dispersion, would require modulating the effective value of $\chi^{(3)}$ over the length of the medium with a modulation period of $29.69\mu$m. This modulations would also decrease the efficiency in the same way that $n^{th}$-order quasi phase matching decreases the efficiency by a factor of $4/(\pi^{2}n^{2})$. 

Where one cannot periodically pole the nonlinear medium to change the sign of $\chi^{(3)}$ one must instead rely on techniques to create a static modulation of intensity over the nonlinear medium or of modulating the refractive index (which modulates $\chi^{(3)}_{eff}$ from Miller's rule). As an example, if we can modulate the refractive index by $0.01$, this will produce approximately a $17\%$ change in $\chi^{(3)}_{eff}$, but will yield a triplet generation rate smaller by a factor of $0.003$, or about a 25dB reduction in the generation rate compared to perfect phase matching. However, alternative media with tunable properties have been proposed (e.g., hybrid photonic crystal waveguides or periodically tapered fibers)\cite{PhysRevA.101.033840,saleh2019modelling} that offer promising efficiencies.

When a nonlinear medium is strongly driven by a pump laser at high frequency, other processes can occur to generate light at the down-converted wavelength that are not due to $\chi^{(3)}$-SPDC. In particular, spontaneous (Stokes) Raman scattering with optical phonons in the medium can in principle create light within the same band as the down-converted light. This is not the only such confounding process that can occur (e.g., other forms of fluorescence independent of Raman scattering). As to whether these processes will be a detriment will depend on their brightness relative to the triplet generation rate. These confounding sources of light are unlikely to generate ``dark" triplet coincidences at intensities measured with single-photon detectors, but they will put an effective cap in input power, given the maximum detection rate for the photon detectors used.

\section{Conclusion}
In this work:
\begin{itemize}
\item we discussed how to quantify genuine tripartite entanglement in spatial and energy-time degrees of freedom using correlations naturally present in nonlinear optics without assuming a pure quantum state.
\item we derived the triphoton state for third-order SPDC as a source of entangled photon triplets, finding both its spatial wavefunction, and its joint spectral amplitude
\item we analyzed the triphoton wavefunction, and showed that our method can quantify a substantial amount of tripartite spatial entanglement for these states
\item we derived the generation rate of photon triplets in third-order degenerate SPDC to explore how bright these sources may be.
\item we proposed an experiment to measure these spatial correlations in a maximally efficient way that cannot overestimate the entanglement present.
\end{itemize}
By developing the methods to better characterize the genuine tripartite entanglement in these rapidly developing sources, we intend to spur motivation toward new approaches to capitalize on high-dimensional genuine tripartite entanglement as a resource in quantum information technology.

\acknowledgments 
 
We gratefully acknowledge support from the Air Force Office of Scientific Research.

The views expressed are those of the authors and do not reflect the official guidance or position of the United States Government, the Department of Defense or of the United States Air Force. The appearance of external hyperlinks does not constitute endorsement by the United States Department of Defense (DoD) of the linked websites, or of the information, products, or services contained therein. The DoD does not exercise any editorial, security, or other control over the information you may find at these locations.

\bibliography{report} 

\begin{thebibliography}{10}

\bibitem{PhysRevLett.80.2245}
Wootters, W.~K., ``Entanglement of formation of an arbitrary state of two
  qubits,'' {\em Phys. Rev. Lett.}~{\bf 80},  2245--2248 (Mar 1998).

\bibitem{PhysRevLett.76.722}
Bennett, C.~H., Brassard, G., Popescu, S., Schumacher, B., Smolin, J.~A., and
  Wootters, W.~K., ``Purification of noisy entanglement and faithful
  teleportation via noisy channels,'' {\em Phys. Rev. Lett.}~{\bf 76},
  722--725 (Jan 1996).

\bibitem{PhysRevA.57.1619}
Vedral, V. and Plenio, M.~B., ``Entanglement measures and purification
  procedures,'' {\em Phys. Rev. A}~{\bf 57},  1619--1633 (Mar 1998).

\bibitem{christandl2004squashed}
Christandl, M. and Winter, A., ``“squashed entanglement”: an additive
  entanglement measure,'' {\em Journal of mathematical physics}~{\bf 45}(3),
  829--840 (2004).

\bibitem{PhysRevA.65.032314}
Vidal, G. and Werner, R.~F., ``Computable measure of entanglement,'' {\em Phys.
  Rev. A}~{\bf 65},  032314 (Feb 2002).

\bibitem{RevModPhys.81.865}
Horodecki, R., Horodecki, P., Horodecki, M., and Horodecki, K., ``Quantum
  entanglement,'' {\em Rev. Mod. Phys.}~{\bf 81},  865--942 (Jun 2009).

\bibitem{SchneelochQuantTripCVEnt2022}
Schneeloch, J., Birrittella, R.~J., Tison, C.~C., Howland, G.~A., Fanto, M.~L.,
  and Alsing, P.~M., ``Quantifying tripartite spatial and energy-time
  entanglement in nonlinear optics,'' {\em Phys. Rev. Res.}~{\bf 4},  023153
  (May 2022).

\bibitem{Szalay_MultiEntMeas}
Szalay, S., ``Multipartite entanglement measures,'' {\em Phys. Rev. A}~{\bf
  92},  042329 (Oct 2015).

\bibitem{leuchs2019quantum}
Walter, M., Gross, D., and Eisert, J.,  [{\em Quantum Information: From
  Foundations to Quantum Technology Applications}{\nolinebreak\hspace{0.1em}]},
  ch.~14,  305, Wiley-VCH Verlag GmbH {\&} Co. (2019).

\bibitem{guhne2009entanglement}
G{\"u}hne, O. and T{\'o}th, G., ``Entanglement detection,'' {\em Physics
  Reports}~{\bf 474}(1-6),  1--75 (2009).

\bibitem{PhysRevA.83.062321}
Huber, M. and Plesch, M., ``Purification of genuine multipartite
  entanglement,'' {\em Phys. Rev. A}~{\bf 83},  062321 (Jun 2011).

\bibitem{ModiDiscordRMP2012}
Modi, K., Brodutch, A., Cable, H., Paterek, T., and Vedral, V., ``The
  classical-quantum boundary for correlations: Discord and related measures,''
  {\em Rev. Mod. Phys.}~{\bf 84},  1655--1707 (Nov 2012).

\bibitem{Schneeloch_TriEnt}
Schneeloch, J., Tison, C.~C., Fanto, M.~L., Ray, S., and Alsing, P.~M.,
  ``Quantifying tripartite entanglement with entropic correlations,'' {\em
  Phys. Rev. Research}~{\bf 2},  043152 (Oct 2020).

\bibitem{Berta2010}
Berta, M., Christandl, M., Colbeck, R., Renes, J.~M., and Renner, R., ``The
  uncertainty principle in the presence of quantum memory,'' {\em Nature
  Physics}~{\bf 6}(9),  659 (2010).

\bibitem{SchneelochPra2018}
Schneeloch, J. and Howland, G.~A., ``{Quantifying high-dimensional entanglement
  with Einstein-Podolsky-Rosen correlations},'' {\em Phys. Rev. A}~{\bf 97},
  042338 (Sep 2018).

\bibitem{Cover2006}
Cover, T.~M. and Thomas, J.~A.,  [{\em {Elements of Information
  Theory}}{\nolinebreak\hspace{0.1em}]}, Wiley and Sons, New York, second~ed.
  (2006).

\bibitem{boyd2007nonlinear}
Boyd, R.~W.,  [{\em Nonlinear optics}{\nolinebreak\hspace{0.1em}]}, Academic
  press, third~ed. (2007).

\bibitem{Schneeloch_SPDC_Reference_2016}
Schneeloch, J. and Howell, J.~C., ``Introduction to the transverse spatial
  correlations in spontaneous parametric down-conversion through the biphoton
  birth zone,'' {\em Journal of Optics}~{\bf 18}(5),  053501 (2016).

\bibitem{EPR1935}
Einstein, A., Podolsky, B., and Rosen, N., ``{Can Quantum-Mechanical
  Description of Physical Reality Be Considered Complete?},'' {\em Phys.
  Rev.}~{\bf 47},  777--780 (May 1935).

\bibitem{PhysRevLett.88.120401}
Mancini, S., Giovannetti, V., Vitali, D., and Tombesi, P., ``Entangling
  macroscopic oscillators exploiting radiation pressure,'' {\em Phys. Rev.
  Lett.}~{\bf 88},  120401 (Mar 2002).

\bibitem{tomov1992third}
Tomov, I., Van~Wonterghem, B., and Rentzepis, P.~M., ``Third-harmonic
  generation in barium borate,'' {\em Applied optics}~{\bf 31}(21),  4172--4174
  (1992).

\bibitem{miyata2011high}
Miyata, K., Petrov, V., and Noack, F., ``High-efficiency single-crystal
  third-harmonic generation in bib3o6,'' {\em Optics letters}~{\bf 36}(18),
  3627--3629 (2011).

\bibitem{gayer2008temperature}
Gayer, O., Sacks, Z., Galun, E., and Arie, A., ``Temperature and wavelength
  dependent refractive index equations for mgo-doped congruent and
  stoichiometric linbo3,'' {\em Applied Physics B}~{\bf 91},  343--348 (2008).

\bibitem{schneeloch2018EntExp}
Schneeloch, J., Tison, C.~C., Fanto, M.~L., Alsing, P.~M., and Howland, G.~A.,
  ``Quantifying entanglement in a 68-billion-dimensional quantum state space,''
  {\em Nat. Commun.}~{\bf 10}(2785) (2019).

\bibitem{hamel2014direct}
Hamel, D.~R., Shalm, L.~K., H{\"u}bel, H., Miller, A.~J., Marsili, F., Verma,
  V.~B., Mirin, R.~P., Nam, S.~W., Resch, K.~J., and Jennewein, T., ``Direct
  generation of three-photon polarization entanglement,'' {\em Nature
  Photonics}~{\bf 8}(10),  801--807 (2014).

\bibitem{PhysRevA.101.033840}
Cavanna, A., Hammer, J., Okoth, C., Ortiz-Ricardo, E., Cruz-Ramirez, H.,
  Garay-Palmett, K., U'Ren, A.~B., Frosz, M.~H., Jiang, X., Joly, N.~Y., and
  Chekhova, M.~V., ``Progress toward third-order parametric down-conversion in
  optical fibers,'' {\em Phys. Rev. A}~{\bf 101},  033840 (Mar 2020).

\bibitem{Schneeloch_2019_RateRef}
Schneeloch, J., Knarr, S.~H., Bogorin, D.~F., Levangie, M.~L., Tison, C.~C.,
  Frank, R., Howland, G.~A., Fanto, M.~L., and Alsing, P.~M., ``Introduction to
  the absolute brightness and number statistics in spontaneous parametric
  down-conversion,'' {\em Journal of Optics}~{\bf 21},  043501 (feb 2019).

\bibitem{malitson1965interspecimen}
Malitson, I.~H., ``Interspecimen comparison of the refractive index of fused
  silica,'' {\em Josa}~{\bf 55}(10),  1205--1209 (1965).

\bibitem{PhysRevB.61.10702}
Gubler, U. and Bosshard, C., ``Optical third-harmonic generation of fused
  silica in gas atmosphere: Absolute value of the third-order nonlinear optical
  susceptibility ${\ensuremath{\chi}}^{(3)}$,'' {\em Phys. Rev. B}~{\bf 61},
  10702--10710 (Apr 2000).

\bibitem{saleh2019modelling}
Saleh, M.~F., ``Modelling spontaneous four-wave mixing in periodically tapered
  waveguides,'' {\em Optics Express}~{\bf 27}(9),  11979--11990 (2019).

\bibitem{schneeloch2023negativity}
Schneeloch, J., Shelton~Jacinto, H., Tison, C.~C., and Alsing, P.~M.,
  ``Negativity vs. purity and entropy in witnessing entanglement,'' {\em
  Scientific Reports}~{\bf 13},  4601 (2023).

\bibitem{nielsen2000}
Nielsen, M.~A. and Chuang, I.~L.,  [{\em {Quantum Computation and Quantum
  Information}}{\nolinebreak\hspace{0.1em}]}, Cambridge University Press
  (2000).

\end{thebibliography}
\bibliographystyle{spiebib} 

\newpage
\appendix

\section{Proof of entanglement correlation relation}\label{AppCorRel}
A reduced form of the entanglement correlation relation was proved in \cite{schneeloch2023negativity}:
\begin{equation}\label{weakerRel}
H(Q_{A}:Q_{B})\leq E_{F}(AB) + S(AB)
\end{equation}
Here, we show how this relation is improved to the form:
\begin{equation}
H(Q_{A}:Q_{B})\leq E_{F}(AB) + \min\{S(AB),S(A),S(B)\}
\end{equation}
which is tighter in the case of separable, and presumably nearly separable states.

The monotonicity of the relative entropy \cite{nielsen2000} allows us to assert that the classical correlations quantified by the Shannon mutual information will always be less than or equal to the quantum mutual information.
\begin{equation}\label{monotonicity}
H(Q_{A}:Q_{B})\leq I(A:B)
\end{equation}
This is because the classical mutual information can be interpreted as a quantum mutual information after measurements have been performed on the system.

The quantum mutual information itself can be expressed in terms of marginal and conditional entropies:
\begin{equation}
I(A:B)=S(A)-S(A|B) = S(B)-S(B|A)
\end{equation}

Next, we have that both $-S(A|B)$ and $-S(B|A)$ are lower bounds to the entanglement of formation $E_{F}(AB)$. Together this gives us the two equations:
\begin{subequations}
\begin{align}
I(A:B)&\leq S(A) + E_{F}(AB)\\
I(A:B)&\leq S(B) + E_{F}(AB)
\end{align}
\end{subequations}
Combining these relations with equations \eqref{monotonicity} and \eqref{weakerRel} gives us the three relations:
\begin{subequations}
\begin{align}
H(Q_{A}:Q_{B})&\leq E_{F}(AB)+S(AB)\\
H(Q_{A}:Q_{B})&\leq E_{F}(AB)+S(A)\\
H(Q_{A}:Q_{B})&\leq E_{F}(AB)+S(B)
\end{align}
\end{subequations}
which are readily consolidated into the entanglement-correlation relation used in this paper:
\begin{equation}
H(Q_{A}:Q_{B})\leq E_{F}(AB) + \min\{S(AB),S(A),S(B)\}
\end{equation}

\section{Derivation of triphoton wavefunction for third-order SPDC}\label{AppTriphotDeriv}
In this Appendix, we show how to derive the triphoton spatial wavefunction in third-order $\chi^{(3)}$ SPDC (abbreviated in subscripts as 3SPDC). To start, we have the hamiltonian for the $\chi^{(3)}$SPDC process:
\begin{equation}
\hat{H}_{3SPDC}=\sum_{k_{p},k_{1},k_{2},k_{3}}G_{k_{p},k_{1}k_{2},k_{3}}\hat{a}_{k_{p}}\hat{a}_{k_{1}}^{\dagger}\hat{a}_{k_{2}}^{\dagger}\hat{a}_{k_{3}}^{\dagger} + H.c.
\end{equation}
To obtain the state of the down-converted light, we use first-order time-dependent perturbation theory:
\begin{equation}
|\psi\rangle_{3SPDC}\approx\sum_{k_{p},k_{1},k_{2},k_{3}}G_{k_{p},k_{1}k_{2},k_{3}}\hat{a}_{k_{p}}\hat{a}_{k_{1}}^{\dagger}\hat{a}_{k_{2}}^{\dagger}\hat{a}_{k_{3}}^{\dagger}|vac\rangle
\end{equation}
where factors of $i$ and $\hbar$ have been absorbed into $G$. Since we are only interested in the triphoton wavefunction and not in the absolute probability of triphoton generation here, many factors will be absorbed into an overall normalization constant.

The spatially varying components of $G_{k_{p},k_{1}k_{2},k_{3}}$ are given by:
\begin{equation}
G_{k_{p},k_{1}k_{2},k_{3}}\propto \int d^{3}r\Big[\chi^{(3)}_{eff}(\vec{r})e^{i\Delta\vec{k}\cdot\vec{r}}\Big]
\end{equation}
where $\Delta \vec{k}\equiv \vec{k}_{1}+\vec{k}_{2}+\vec{k}_{3}-\vec{k}_{p}$, and $\chi^{(3)}_{eff}(\vec{r})$ is the spatially varying effective third order susceptibility, regarded as constant within the nonlinear medium, and zero outside.

With these approximations, we can begin to express the triphoton state in terms of a triphoton wavefunction:
\begin{subequations}
\begin{align}
|\psi\rangle_{3SPDC}&\propto\sum_{k_{p},k_{1},k_{2},k_{3}}\Psi(\vec{k}_{1},\vec{k}_{2},\vec{k}_{3})\hat{a}_{k_{1}}^{\dagger}\hat{a}_{k_{2}}^{\dagger}\hat{a}_{k_{3}}^{\dagger}|vac\rangle\\
\Psi(\vec{k}_{1},\vec{k}_{2},\vec{k}_{3})&=\mathcal{N} \sum_{k_{p}}\alpha_{k_{p}}\bigg(\prod_{i=x,y,z}\text{sinc}\bigg(\frac{\Delta k_{i}L_{i}}{2}\bigg)\bigg)
\end{align}
\end{subequations}
where $\alpha_{k_{p}}$ is a coherent state pump amplitude substituting for $\hat{a}_{k_{p}}$ in the undepleted (classically bright) pump approximation.

Next, we assume the pump is sufficiently narrowband in frequency that its longitudinal momentum takes on one value in the sum over $k_{p}$. In addition we take the small-angle/ paraxial approximation so that the pump amplitude $\alpha_{k_{p}}$ factors out as a product of a longitudinal amplitude $\alpha_{pz}$ and a transverse amplitude $\alpha_{qp}$. Together, this gives:
\begin{align}
\Psi(\vec{k}_{1},\vec{k}_{2},\vec{k}_{3})&=\mathcal{N} \int dk_{px}dk_{py}\alpha_{qp}(\vec{q}_{p})\\
&\times \prod_{i=x,y,z}\text{sinc }\bigg(\frac{(k_{1i}+k_{2i}+k_{3i}-k_{pi})L_{i}}{2}\bigg)
\end{align}

Next, we assume the transverse dimensions $L_{x},L_{y}$ of the medium are large enough to completely envelop the transverse profile of the pump and down-converted light, which in turn, is much larger than the pump wavelength. Because $(L_{x},L_{y})\gg\lambda_{p}$, we may treat the transverse sinc functions as delta functions, when integrating over the transverse pump momentum, which also enforces transverse momentum conservation.
\begin{align}
\Psi(\vec{k}_{1},\vec{k}_{2},\vec{k}_{3})&=\mathcal{N} \int dk_{px}dk_{py}\alpha_{qp}(\vec{q}_{p})\\
&\times \text{sinc }\bigg(\frac{(k_{1z}+k_{2z}+k_{3z}-k_{pz})L_{z}}{2}\bigg)\nn\\
&\times\prod_{i=x,y}\delta(k_{1i}+k_{2i}+k_{3i}-k_{pi})
\end{align}

To isolate the transverse spatial component of the triphoton wavefunction, we express the longitudinal momentum components in the sinc function in terms of the respective momentum magnitudes and the transverse spatial components through the Pythagorean formula:
\begin{equation}
k_{z}=\sqrt{|\vec{k}|^{2}-|\vec{q}|^{2}}\approx |\vec{k}| -\frac{|\vec{q}|^{2}}{2|\vec{k}|}
\end{equation}
The approximation comes from the small-angle approximation in which the magnitude of the transverse projection $\vec{q}$ of the momentum $\vec{k}$ is much smaller than the total magnitude of $\vec{k}$. Initially, this complicates the sinc function:
\begin{align}
\Psi(\vec{k}_{1},\vec{k}_{2},\vec{k}_{3})&=\mathcal{N} \int dk_{px}dk_{py}\alpha_{qp}(\vec{q}_{p})\\
&\times \text{sinc }\bigg(\frac{(|\vec{k}_{1}|+|\vec{k}_{2}|+|\vec{k}_{3}|-|\vec{k}_{p}|)L_{z}}{2}\nn\\
&\qquad\qquad-\bigg(\frac{|\vec{q}_{1}|^{2}}{|\vec{k}_{1}|}+\frac{|\vec{q}_{2}|^{2}}{|\vec{k}_{2}|}+\frac{|\vec{q}_{3}|^{2}}{|\vec{k}_{3}|}-\frac{|\vec{q}_{p}|^{2}}{|\vec{k}_{p}|}\bigg)\frac{L_{z}}{4}\bigg)\nn\\
&\times\prod_{i=x,y}\delta(k_{1i}+k_{2i}+k_{3i}-k_{pi})
\end{align}

Following this, we assume the use of narrowband frequency filters to fix the magnitudes of $|\vec{k}_{1}|$, $|\vec{k}_{3}|$, and $|\vec{k}_{4}|$, where $|\vec{k}_{p}|$ has already been fixed in the narrowband pump assumption made earlier. Through optimum phase matching, we can achieve the condition:
\begin{equation}
|\vec{k}_{1}|+|\vec{k}_{2}|+|\vec{k}_{3}|-|\vec{k}_{p}|\approx 0
\end{equation}
greatly simplifying the triphoton wavefunction.

By assuming the pump beam is well-collimated, its transverse momentum amplitude $\alpha(\vec{q}_{p})$ centered at $\vec{q}_{p}=0$ will be narrow enough that we can neglect the sinc function's dependence on $\vec{q}_{p}$:
\begin{align}
\Psi(\vec{k}_{1},\vec{k}_{2},\vec{k}_{3})&=\mathcal{N} \int dk_{px}dk_{py}\alpha_{qp}(\vec{q}_{p})\\
&\times \text{sinc }\bigg(\bigg(\frac{|\vec{q}_{1}|^{2}}{|\vec{k}_{1}|}+\frac{|\vec{q}_{2}|^{2}}{|\vec{k}_{2}|}+\frac{|\vec{q}_{3}|^{2}}{|\vec{k}_{3}|}\bigg)\frac{L_{z}}{4}\bigg)\nn\\
&\times\prod_{i=x,y}\delta(k_{1i}+k_{2i}+k_{3i}-k_{pi})
\end{align}
 At this point, we can perform the integration over transverse pump momenta to eliminate the delta functions:
\begin{align}
\Psi(\vec{k}_{1},\vec{k}_{2},\vec{k}_{3})&\approx\mathcal{N} \alpha_{qp}(\vec{q}_{1}+\vec{q}_{2}+\vec{q}_{3})\\
&\times \text{sinc }\bigg(\bigg(\frac{|\vec{q}_{2}+\vec{q}_{3}-\vec{q}_{p}|^{2}}{|\vec{k}_{1}|}\nn\\
&+\frac{|\vec{q}_{1}+\vec{q}_{3}-\vec{q}_{p}|^{2}}{|\vec{k}_{2}|}+\frac{|\vec{q}_{1}+\vec{q}_{2}-\vec{q}_{p}|^{2}}{|\vec{k}_{3}|}\bigg)\frac{L_{z}}{4}\bigg)\nn
\end{align}
To simplify this wavefunction further, we re-apply the assumption that the pump has sufficiently narrow transverse momentum bandwidth that dependence on transverse pump momentum can be eliminated from the sinc function:
\begin{align}
\Psi(&\vec{k}_{1},\vec{k}_{2},\vec{k}_{3})\approx\mathcal{N} \alpha_{qp}(\vec{q}_{1}+\vec{q}_{2}+\vec{q}_{3})\\
&\times \text{sinc }\bigg(\bigg(\frac{|\vec{q}_{2}+\vec{q}_{3}|^{2}}{|\vec{k}_{1}|}+\frac{|\vec{q}_{1}+\vec{q}_{3}|^{2}}{|\vec{k}_{2}|}+\frac{|\vec{q}_{1}+\vec{q}_{2}|^{2}}{|\vec{k}_{3}|}\bigg)\frac{L_{z}}{4}\bigg)\nn
\end{align}
For our triphoton spatial wavefunction, we assume $|\vec{k}_{1}|=|\vec{k}_{2}|=|\vec{k}_{3}|=|\vec{k}_{p}|/3$, which gives us the final simplified form:
\begin{align}
\Psi(&\vec{k}_{1},\vec{k}_{2},\vec{k}_{3})\approx\mathcal{N} \alpha_{qp}(\vec{q}_{1}+\vec{q}_{2}+\vec{q}_{3})\\
&\times \text{sinc }\bigg(\Big(|\vec{q}_{2}+\vec{q}_{3}|^{2}+|\vec{q}_{1}+\vec{q}_{3}|^{2}+|\vec{q}_{1}+\vec{q}_{2}|^{2}\Big)\frac{L_{z}}{4|\vec{k}_{1}|}\bigg)\nn
\end{align}
Where this wavefunction is approximately separable into horizontal and vertical spatial components, we have the 1D triphoton wavefunction in rotated coordinates:
\begin{equation}
\psi(k_{u},k_{v},k_{w})\approx\mathcal{N} \alpha_{p}(k_{u}\sqrt{3}) \text{sinc }\bigg(\frac{L_{z}}{4|\vec{k}_{1}|}\Big(4 k_{u}^{2} + k_{v}^{2} + k_{w}^{2}\Big)\bigg)
\end{equation}
Among other things, we see that the triphoton wavefunction for a gaussian pump beam is well-approximated by a triple-gaussian wavefunction.

\section{Derivation of triphoton generation rate for third-order SPDC}\label{AppTriPhotBrightness}

From \cite{Schneeloch_2019_RateRef}, the nonlinear Hamiltonian for third-order Spontaneous Parametric Down-Conversion (SPDC) is given by:
\begin{equation}
\hat{H}_{NL}=\frac{1}{4}\int d^{3}r\;\big(\zeta_{ij\ell m}^{(3)}(\vec{r})\hat{D}_{i}^{+}(\vec{r},t)\hat{D}_{j}^{-}(\vec{r},t)\hat{D}_{\ell}^{-}(\vec{r},t)\hat{D}_{m}^{-}(\vec{r},t) + H.c.,\big),
\end{equation}
where repeated coordinate indices are summed over according to the Einstein summation convention.  Here: $\zeta_{ij\ell m}^{(3)}(\vec{r})$ is the third-order inverse optical susceptibility tensor of the medium the light is interacting with, and $\hat{D}_{i}^{+}(\vec{r},t)$ is the positive frequency component of the electric displacement field operator, such that in the basis of propagating (Hermite) Gaussian beam modes we have:
\begin{equation}
\hat{D}^{+}(\vec{r},t)=i\!\!\sum_{\vec{\mu},k_{z},s}\!\!\sqrt{\frac{\epsilon_{0}n^{2}_{k_{z}}\!\hbar\omega_{k_{z}}}{2 \mathcal{L}_{z}}}\vec{\epsilon}_{k_{z},s}g_{\vec{\mu}}(x,y)e^{i k_{z} z}e^{-i\omega t}\hat{a}_{\vec{\mu},k_{z},s}.
\end{equation}
In this expression: $\epsilon_{0}$ is the permittivity of free space; $n_{\vec{k}}$ is the index of refraction for a photon of momentum $\vec{p}=\hbar \vec{k}$, where we use $\vec{k}$ to index momentum to eliminate extra factors of $\hbar=h/2\pi$, the quantum of angular momentum. With this, we point out that the angular frequency of the photon $\omega_{k}$ is also indexed by its momentum. In addition: $\hat{a}_{\vec{\mu},k_{z},s}$ is the annihilation operator of a photon with transverse spatial mode function $g_{\vec{\mu}}(x,y)$ indexed by $\vec{\mu}$, longitudinal momentum indexed by $k_{z}$ and polarization indexed by $s$, and $\vec{\epsilon}_{k_{z},s}$ is a unit vector in the direction of that polarization. The field modes are initially defined with respect to a rectangular cavity with volume $V=\mathcal{L}_{x}\mathcal{L}_{y}\mathcal{L}_{z}$, but the transverse cavity dimensions disappear upon converting to the basis of transverse gaussian modes. To distinguish the dimensions of the quantization volume from the dimensions of the nonlinear medium, we let $\{L_{x},L_{y},L_{z}\}$ be the dimensions of the nonlinear medium.

In this nonlinear-optical Hamiltonian, we are isolating the process of third-order SPDC by considering only terms where pump photons are annihilated and photon triplets are created (and vise versa). A fuller treatment including all other interactions of the nonlinear medium with a bright undepleted pump will include the intensity-dependent refractive index (i..e, self-phase modulation) which effects the phase-matching properties of third-order down-conversion as well. However, we may regard this effect as accounted for in the refractive indices of the medium used to predict the overall brightness of photon triplet generation. Knowing this, we proceed with the rest of the calculation.

Next, we assume we are using one polarization of the pump, signal, and idler fields, and carry out the summation over the components of the inverse nonlinear susceptibility $\zeta^{(3)}$. We also define $\Delta k_{z} \equiv k_{1z}+k_{2z}+k_{3z}-k_{pz}$ and $\Delta\omega \equiv \omega_{1}+\omega_{2}+\omega_{3}-\omega_{p}$. Finally, we assume the pump is bright enough to be treated as an undepleted classical field, and replace its annihilation operator with a corresponding coherent state amplitude:
\begin{align}
\hat{H}_{NL}&=6\sum_{k_{pz}}\sum_{k_{1z}}\sum_{k_{2z}}\sum_{k_{3z}}\chi_{eff}^{(3)}\sqrt{\frac{\hbar^{4}\omega_{p}\omega_{1}\omega_{2}\omega_{3}}{16 \epsilon_{0}^{2} \mathcal{L}_{z}^{4}n^{2}_{p}n^{2}_{1}n^{2}_{2}n_{3}^{2}}}\nn\\
&\qquad\cdot \int d^{3}r\;\big(\bar{\chi}_{eff}^{(3)}(\vec{r})g_{\vec{\mu}_{p}}(x,y)g_{\vec{\mu}_{1}}^{*}(x,y)g_{\vec{\mu}_{2}}^{*}(x,y)g_{\vec{\mu}_{3}}^{*}(x,y)e^{-i\Delta k_{z} z}\big)\nn\\
&\qquad\qquad\cdot e^{i\Delta\omega t}\cdot \alpha_{\vec{\mu}_{p},k_{pz}}\hat{a}_{\vec{\mu}_{1},k_{1z}}^{\dagger}\hat{a}_{\vec{\mu}_{2},k_{2z}}^{\dagger}\hat{a}_{\vec{\mu}_{3},k_{3z}}^{\dagger}+ H.c.
\end{align}
where here, we use the approximate relation between forward and inverse third-order susceptibilities in a centrosymmetric medium:
\begin{equation}\label{Inversechi3Formula}
\zeta_{eff}^{(3)}=-\frac{\chi_{eff}^{(3)}}{\epsilon_{0}^{3}n_{p}^{2}n_{1}^{2}n_{2}^{2}n_{3}^{2}}.
\end{equation}
and introduce the nonlinearity profile $\bar{\chi}_{eff}^{(3)}$ that is zero outside the medium and unity (or otherwise modulated) inside the medium (so that $\chi^{(3)}_{eff}(z)=\chi^{(3)}_{eff}\bar{\chi}^{(3)}_{eff}(z)$).

Approximating the Hamiltonian as a continuous integral over frequency instead of a discrete sum of momenta by the relations:
\begin{subequations}
\begin{align}
\sum_{k}\hat{a}_{k}^{\dagger}\hat{a}_{k} &\approx \int dk \hat{a}^{\dagger}(k)\hat{a}(k)\approx \int d\omega \hat{a}^{\dagger}(\omega)\hat{a}(\omega).\\
\hat{a}^{\dagger}(\omega) &\approx \sqrt{\frac{\mathcal{L}_{z}n_{g}}{2\pi c}} \hat{a}_{k}^{\dagger}\\
\sum_{k_{1z}}&\approx\Big(\frac{\mathcal{L}_{z}}{2\pi}\Big)\frac{n_{g1}}{c}\int d\omega_{1},
\end{align}
\end{subequations}
and assuming that the portion of the Hamiltonian $\sqrt{\omega_{p}\omega_{1}\omega_{2}\omega_{3}}$ is slowly varying enough over the frequency bands considered that it may be taken as a constant (with their central values) outside the integral, we can obtain the simplified Hamiltonian:
\begin{align}
\hat{H}_{NL}\!\!&=\!\hbar\! \left(\!\!\frac{3\hbar \chi_{eff}^{(3)}}{8 \pi^{2} \epsilon_{0} c^{2}}\!\!\right)\!\!\sqrt{\frac{n_{g1}n_{g2}n_{g3}n_{gp}}{n^{2}_{p}n^{2}_{1}n^{2}_{2}n_{3}^{2}}}\!\!\sqrt{\omega_{p0}\omega_{10}\omega_{20}\omega_{30}}\!\!\!\int\!\! d\omega_{p}d\omega_{1}d\omega_{2}d\omega_{3}\alpha(\omega_{p})\Phi(\Delta k_{z})e^{i\Delta\omega t}\!\!\cdot\! \hat{a}^{\dagger}_{\vec{\mu}_{1}}\!(\omega_{1})\hat{a}^{\dagger}_{\vec{\mu}_{2}}\!(\omega_{2})\hat{a}^{\dagger}_{\vec{\mu}_{3}}\!(\omega_{3})\!+\! h.c.
\end{align}
where
\begin{equation}
\Phi(\Delta k_{z})\equiv\int d^{3}r\;\big(\bar{\chi}_{eff}^{(3)}(\vec{r})g_{\vec{\mu}_{p}}(x,y)g_{\vec{\mu}_{1}}^{*}(x,y)g_{\vec{\mu}_{2}}^{*}(x,y)g_{\vec{\mu}_{3}}^{*}(x,y)e^{-i\Delta k_{z} z}\big),
\end{equation}
is an overlap integral decribing the phase-matching of this process.

To find the state of the down-converted triplet, we assume a pump pulse with an average $N_{p}$ photons, so that $\alpha(\omega_{p})=s(\omega_{p})\sqrt{N_{P}}$ (and $|s(\omega_{p})|^{2}$ integrates to unity), and use first-order time-dependent perturbation theory:
\begin{subequations}
\begin{align}
|\Psi_{final}\rangle &\approx \bigg( \mathbf{I} -\frac{i}{\hbar}\int_{-\infty}^{\infty} \!\!dt' \hat{H}_{NL}(t') \bigg)|\alpha_{p},0,0,0\rangle\\
&= |\alpha_{p},0,0,0\rangle + |\alpha_{p},\Psi_{triplet}\rangle,
\end{align}
\end{subequations}
Where $e^{i\Delta\omega t}$ is the only time-dependent part of the Hamiltonian, integrating over all time produces a Dirac Delta function $\delta(\Delta\omega)$, ensuring energy conservation. If we trace over the pump field, which is independent under our assumptions, we obtain for the state of the down-converted triplet:
\begin{equation}
|\Psi_{triplet}\rangle =\int d\omega_{1}d\omega_{2}d\omega_{3} \;\psi(\omega_{1},\omega_{2},\omega_{3})\cdot \hat{a}^{\dagger}_{\vec{\mu}_{1}}(\omega_{1})\hat{a}^{\dagger}_{\vec{\mu}_{2}}(\omega_{2})\hat{a}^{\dagger}_{\vec{\mu}_{3}}(\omega_{3})|0,0,0\rangle.
\end{equation}
where $\psi(\omega_{1},\omega_{2},\omega_{3})$ is the joint spectral amplitude of the photon triplet:
\begin{equation}
\psi(\omega_{1},\omega_{2},\omega_{3})=-i\sqrt{N_{p}}\left(\frac{3\hbar \chi_{eff}^{(3)}}{4 \pi \epsilon_{0} c^{2}}\right)\sqrt{\frac{n_{g1}n_{g2}n_{g3}n_{gp}}{n^{2}_{p}n^{2}_{1}n^{2}_{2}n_{3}^{2}}}\sqrt{\omega_{p0}\omega_{10}\omega_{20}\omega_{30}}s(\omega_{1}+\omega_{2}+\omega_{3})\Phi(\Delta k_{z}),
\end{equation}
and the overall number of photon triplets $\langle N_{triplets}\rangle$ can be found through integrating the joint spectral intensity:
\begin{equation}
\langle N_{triplets}\rangle = \int d\omega_{1}d\omega_{2}d\omega_{3}|\psi(\omega_{1},\omega_{2},\omega_{3})|^{2}.
\end{equation}

To simplify the calculation of $\langle N_{triplets}\rangle$, we focus on the overlap integral $\Phi(\Delta k_{z})$. Assuming that we are using only zero-order Gaussian beam modes, and that $\sigma_{1}=\sigma_{2}=\sigma_{3}=\sigma_{p}\sqrt{3}$, the overlap integral simplifies to:
\begin{equation}
\Phi(\Delta k_{z})=\frac{L_{z}}{\sqrt{108}\pi\sigma_{p}^{2}}\text{Sinc}\Big(\frac{\Delta k_{z} L_{z}}{2}\Big)
\end{equation}
Expressing $\Delta k_{z}$ in terms of frequency is best done in shifted rotated frequency coordinates:
\begin{subequations}
\begin{align}
\delta\omega_{u}&=\sqrt{\frac{1}{3}}(\delta\omega_{1}+\delta\omega_{2}+\delta\omega_{3})\\
\delta\omega_{v}&=-\sqrt{\frac{2}{3}}\delta\omega_{1} + \sqrt{\frac{1}{6}}\delta\omega_{2} + \sqrt{\frac{1}{6}}\delta\omega_{3}\\
\delta\omega_{w}&=\sqrt{\frac{1}{2}}(\delta\omega_{2}-\delta\omega_{3})
\end{align}
\end{subequations}
where for example, $\delta \omega_{1}= (\omega_{1}-\omega_{10})$.

Assuming $\omega_{10}=\omega_{20}=\omega_{30}=\omega_{p0}/3$ and a narrow enough bandwidth pump that we may eliminate dependence on $\delta\omega_{u}$ from the phase matching function, we obtain the simplified expression for $\Delta k_{z}$:
\begin{equation}
\Delta k_{z} \approx \frac{\kappa_{0}}{2}(\delta\omega_{v}^{2}+\delta\omega_{w}^{2}).
\end{equation}
so that:
\begin{equation}
\langle N_{triplets}\rangle = \frac{N_{p}L_{z}^{2}}{\sigma_{p}^{4}}\frac{\hbar^{2} }{5184\sqrt{3}\pi^{4}\epsilon_{0}^{2}c^{4}}\frac{n_{g1}n_{g2}n_{g3}n_{gp}}{n^{2}_{p}n^{2}_{1}n^{2}_{2}n_{3}^{2}}(\chi_{eff}^{(3)})^{2}\omega_{p0}^{4}\!\!\int \!\!d\delta\omega_{u}d\delta\omega_{v}d\delta\omega_{w}|s(\delta\omega_{u})|^{2}\text{Sinc}^{2}(\frac{\kappa_{0}L_{z}}{4}(\delta\omega_{v}^{2}+\delta\omega_{w}^{2})).
\end{equation}
The integral over $\delta\omega_{u}$ is readily done, and the integrals over $\delta\omega_{v}$ and $\delta\omega_{w}$ are doable in polar coordinates. Ultimately, this leads us to the formula for $\langle N_{triplets}\rangle$:
\begin{equation}
\langle N_{triplets}\rangle = \frac{\hbar^{2} }{2592\sqrt{3}\pi^{2}\epsilon_{0}^{2}c^{4}}\frac{n_{g1}n_{g2}n_{g3}n_{gp}}{n^{2}_{p}n^{2}_{1}n^{2}_{2}n_{3}^{2}}\frac{(\chi_{eff}^{(3)})^{2}\omega_{p0}^{4}}{|\kappa_{0}|}\frac{N_{p}L_{z}}{\sigma_{p}^{4}}.
\end{equation}
To obtain a rate of triplet generation $R_{triplets}$, it is enough to substitute the average pump photon number $N_{p}$ with the pump photon number rate, which can be expressed in terms of pump power $P\approx\hbar\omega_{p0}R_{p}$:
\begin{equation}
\langle R_{triplets}\rangle = \frac{\hbar }{2592\sqrt{3}\pi^{2}\epsilon_{0}^{2}c^{4}}\frac{n_{g1}n_{g2}n_{g3}n_{gp}}{n^{2}_{p}n^{2}_{1}n^{2}_{2}n_{3}^{2}}\frac{(\chi_{eff}^{(3)})^{2}\omega_{p0}^{3}}{|\kappa_{0}|}\frac{P L_{z}}{\sigma_{p}^{4}}.
\end{equation}
Note: Although we assumed a pump pulse for this derivation, a CW pump beam can be interpreted as a continuous mixture of pump pulses. Because of this, the number of photon triplets generated is a corresponding continuous sum of the triplets generated from individual pulses.

\pagestyle{fancy}

\end{document}